\documentclass[11pt]{article}
\usepackage{epsfig}
\usepackage{amsmath}
\usepackage{amssymb}
\usepackage{boxedminipage}

\newcommand {\be}[1]{\begin{equation}\label{#1}}
\newcommand {\ee}{\end{equation}}
\newcommand {\bea}{\begin{eqnarray}}
\newcommand {\eea}{\end{eqnarray}}

\newcommand{\qed}{\hfill \rule{7pt}{7pt}}

\newtheorem{theorem}{Theorem}
\newtheorem{lemma}{Lemma}

\begin{document}

\title{Groupstrategyproofness of the Egalitarian Mechanism for
Constrained
Rationing Problems}

\author{
Shyam Chandramouli\thanks{IEOR Department, Columbia University, New York,
NY; {\tt sc3102@columbia.edu}} and
Jay\
Sethuraman\thanks{IEOR Department, Columbia University, New York,
NY;
{\tt jay@ieor.columbia.edu}}}

\date{June 2011}

\maketitle

\begin{abstract}
Motivated by applications in many economic environments, Bochet et
al.~\cite{bims10} generalize
the classic rationing model (see Sprumont~\cite{spr91})as follows: there is a 
moneyless market, in which a non-storable,
homogeneous commodity is reallocated between agents with single-peaked
preferences.
Agents are either suppliers or demanders. Transfers between a supplier and a
demander are feasible only if they are \textit{linked}, and the links form
an arbitrary bipartite graph. Information about individual preferences is
private, 
and so is information
about feasible links: an agent may unilaterally close one of her links if it is
in 
her interest to do so. For this problem they propose the
 {\em egalitarian transfer} solution, which
 equalizes the net transfers of rationed agents as much
as permitted by the bilateral constraints. Furthermore, they show that the
egalitarian mechanism elicits a truthful report of
\textit{both }preferences and links. In the variant where demanders are not
strategic but
demands need to be exactly met~\cite{bim10}, they propose a similar mechanism
for which
truthfully reporting the peaks is a dominant strategy, but truthful reporting of
links is not.

The key contribution of the paper is a comprehensive study of the egalitarian 
mechanism with respect to manipulation by a coalition of agents.
Our main result is that the egalitarian mechanism is, in fact, {\em peak group
strategyproof}
: no coalition of agents can (weakly)  benefit from jointly
misreporting their
peaks. Furthermore, we show that the egalitarian mechanism cannot be manipulated
by
any coalition of suppliers (or any coalition of demanders) in the model where
both the
suppliers and demanders are agents. Our proofs shed light on the structure of
the
two models and simpify some of the earlier proofs of strategyproofness.
An implication of  our results is that the well known algorithm
of
Megiddo~\cite{meg77} to compute a lexicographically optimal flow in a network
is 
group strategyproof with respect to the source capacities and sink capacities.
\end{abstract}

\newpage

\section{Introduction}
Motivated by applications in diverse settings, Bochet et al.~\cite{bims10, bim10}
study a model in which a homogeneous commodity is reallocated between a given set
of agents with single-peaked preferences. In this environment, each agent is
endowed with a certain quantity of the commodity and has an ideal consumption
level (his {\em peak}) of that commodity. An agent who is endowed with more than
his ideal consumption level can thus be thought of as a {\em supplier}, and an
agent who is endowed with less than his ideal consumption level can be thought of
as a {\em demander}. Furthermore, transfers are possible only between certain
pairs of agents, represented by a graph.
The goal is to reallocate the commodity to balance supply
and demand to the extent possible. 
The key
difference from conventional economic models on this topic is the 
inability to use money: motivating applications include assigning
(or reassigning) patients to hospitals, assigning students to schools, and
allocating emergency aid supplies. 
On the other hand, it is easy to see that the resulting problem is 
essentially a transportation problem in a (bipartite) network.
The distinguishing feature here is that
the preferences of the agents (such as their peaks) and the other agents
they are linked to is typically private information, so the agents must be given
an incentive to report this information truthfully. 

Bochet et al.~\cite{bims10} propose a clearinghouse mechanism (a centralized 
organization of the market)
that prescribes an allocation that is efficient with respect to (reported) 
preferences and (reported) feasible links between agents. 
They identify a unique {\em egalitarian} allocation---so named because of the
intimate connection with the egalitarian solution of an associated supermodular
game---that Lorenz dominates all Pareto efficient allocations for this problem.
Furthermore, they show that the egalitarian mechanism is strategyproof with
respect to both links and peaks: no {\em individual} agent can strictly benefit
by misreporting his peak or the set of agents he is linked to. In a companion
paper, Bochet et al.~\cite{bim10} consider a ``one-sided''
 model where the demanders are not
strategic, and their demands have to be met exactly. For this model, they propose
an egalitarian mechanism that is strategyproof with respect to peaks, but not
with respect to links. 

Our main result is that the egalitarian mechanism is {\em group} strategyproof 
with respect to peaks in both the one-sided and two-sided models of Bochet et al.
Furthermore, we show that under the egalitarian mechanism it is a weakly dominant
strategy for {\em any} coalition of suppliers (or any coalition of demanders) to
truthfully report their links. These results thus properly generalize the corresponding
(individual) strategyproofness results of Bochet et al.
Our proofs result in an improved understanding of the two models and 
simpify some of the earlier proofs of strategyproofness.

The models of Bochet et al.~\cite{bims10, bim10} generalize many well-known and
well-understood models in the literature; we briefly discuss those that are very
closely related to our work. If there is a single demander (or a single supplier),
the problem reduces to a classical rationing problem of the sort considered by
Sprumont~\cite{spr91}. The egalitarian rule then reduces to the ``uniform'' rule,
and admits many characterizations. If the peaks are all identically 1, the problem
reduces to a matching problem with dichotomous preferences, discussed in Bogomolnaia
and Moulin~\cite{bm04}: in this case, the flow between a supplier-demander pair
can be thought of as the probability that this pair is matched. Some of the negative
results related to link strategyproofness discussed later are true even in this
restricted setting as has already been observed there; we mention these results in
the appropriate sections for the sake of completeness. Finally,
Megiddo~\cite{meg74,meg77}
considered the problem of finding an ``optimal'' flow in a multiple-source,
multiple-sink network, and proposed an algorithm to find a lexicographically
optimal flow. The egalitarian algorithm described in Bochet 
et al.~\cite{bims10, bim10} is essentially Megiddo's algorithm to compute
a lexicographically optimal flow. An implication of our result is that 
Megiddo's algorithm is group strategyproof with respect to the source and sink
capacities, that is, if the agents are located on the edges incident to 
sources and sinks, and all other edge-capacities are common knowledge, then
no coalition of agents have an incentive to misreport their capacities.
This observation is useful in settings in which equitably sharing resources
is important, such as the {\em sharing} problem of Brown~\cite{brown79}.


\section{Models and Notation}

A single commodity is transferred from a set $S$ of suppliers (typically
indexed by $i$) to a set $D$
of demanders (typically indexed by $j$). The commodity can only be transferred between certain
supplier-demander pairs, given by a bipartite graph $G \subseteq S \times D$:
$(i,j) \in G$ means that supplier $i$ can send the commodity to demander $j$.
A set of transfers from the suppliers to the demanders results in a vector $(x,y)\in \mathbb{R%
}_{+}^{S}\times \mathbb{R}_{+}^{D}$ where $x_{i}$ (resp. $y_{j}$) is
supplier $i$'s (resp. demander $j$'s) \textit{net transfer, }with $%
\sum_{S}x_{i}=\sum_{D}y_{j}$.
Each supplier $i$ has \emph{single-peaked preferences}\footnote{%
Writing $P_{i}$ for agent $i$'s strict preference, we have for every $%
x_{i},x_{i}^{\prime }$: $x_{i}<x_{i}^{\prime }\leq s_{i}$ $\Rightarrow
x_{i}^{\prime }P_{i}x_{i},$ and $s_{i}\leq x_{i}<x_{i}^{\prime }$ $%
\Rightarrow x_{i}P_{i}x_{i}^{\prime }$.} $R_{i}$ (with corresponding
indifference relation $I_{i}$) over her \textit{net transfer} $x_{i}$, with
peak $s_{i}$, and each demander $j$ has single-peaked preferences $R_{j}$ ($%
I_{j}$) over her net transfer $y_{j}$, with peak $d_{j}$. We write $\mathcal{%
R}$ for the set of single peaked preferences over $\mathbb{R}_{+}$, and $%
\mathcal{R}^{S\cup D}$ for the set of preference profiles.

We use the following notation. For any subset $T\subseteq S$,
the set of demanders compatible with the suppliers in $T$ is $%
f(T)=\{j\in D|G(T,\{j\})\neq \varnothing \}$. Similarly, the set of suppliers
compatible with the demanders in $C \subseteq D$ is $g(C)=\{i\in S|G(\{i\},C)\neq
\varnothing \}$. We abuse notation and say $f(i)$ and $g(j)$ instead of
$f(\{i\})$ and $g(\{j\})$ respectively.
For any subsets $T \subseteq S$, $C \subseteq D$, $x_T :=
\sum_{i \in T} x_i$ and $y_C := \sum_{j \in C} y_j$.
A transfer of the commodity  from $S$ to $D$ is realized by a flow $\varphi $,
i.e., a vector $\varphi \in \mathbb{R}_{+}^{G}$. We write $x(\varphi
),y(\varphi )$ for the transfers implemented by $\varphi $, namely:%
\begin{equation}
\text{for all }i\in S:x_{i}(\varphi )=\sum_{j\in f(i)}\varphi _{ij};\text{
for all }j\in D:y_{j}(\varphi )=\sum_{i\in g(j)}\varphi _{ij}  \label{1}
\end{equation}%
We say that the net transfers $(x,y)$ are \textit{feasible} if they are
implemented by some flow. Given the graph $G$ and the preferences of the
agents (both suppliers and demanders), we would like to find feasible
net transfers satisfying some desirable properties. An allocation rule
(or a mechansim) is a function that associates feasible net transfers
to any given problem. We shall be concerned exclusively with the
egalitarian mechanism (described later) proposed by 
Bochet et al.~\cite{bims10, bim10}. An important feature of the egalitarian
mechanism is that it is {\em peak-only}: the egalitarian net transfers
depend on the preferences of the agents {\em only} through the peaks. Thus
it makes sense to talk of the {\em problem} $(G,s,d)$: this emphasizes the
fact that the {\em peaks} of the agents and the {\em identity of their
potential trading partners on the other side} can both be modeled as
private information. To summarize: (i) agents (suppliers and demanders) 
report their peaks as well as the set of agents on the other side they are 
compatible with; (ii) the graph $G$ has a link from supplier $i$ to demander $j$
if and only if both $i$ and $j$ report each other as compatible; and (iii) the
egaliatarian mechanism is applied to the problem $(G,s,d)$ where $s$ and
$d$ are the reported peaks of the suppliers and demanders respectively.

A mechanism is {\em link strategyproof} if for any  profile of peaks $(s,d)$,
it is a weakly dominant strategy for each agent to truthfully report their set of compatible
partners. A mechansim is said to be {\em peak strategyproof} if for any
graph $G$ it is a weakly dominant strategy for an agent to truthfully report
their peak.
A mechanism is said to be {\em strategyproof} if it is a weakly dominant strategy
for each agent to truthfully report their peak as well their set of compatible
partners. It is not difficult to see that a mechanism is strategyproof
if and only if it is both peak strategyproof and link strategyproof. These
definitions admit a natural extension that models potential deviations by
coalitions of agents: thus, a mechanism is {\em link group strategyproof} if for
any profile of peaks $(s,d)$ it is a weakly dominant strategy for any coalition of agents
to truthfully report their set of compatible partners. Similarly, a mechanism is {\em peak
group strategyproof} if for any graph $G$ it is a weakly dominant strategy for any
coalition of agents to truthfully report their peaks. Finally, a mechanism is {\em group
strategyproof} if it is a weakly dominant strategy for any coalition of agents to
truthfully report both their peaks and compatible partners. Again, it is not hard to
see that a mechanism is group strategyproof  if it is both peak group strategyproof 
and link group strategyproof.

We shall focus on two related, but distinct, models. In the one-sided model,
only the suppliers are modeled as strategic agents; the demanders are
not strategic and their demands must be met exactly, so that some suppliers
may be forced to send more than their peaks. In the two-sided model,
both the suppliers and the demanders are modeled as strategic agents. Thus, in 
considering strategic issues in the one-sided model, we shall naturally only examine 
coalitions of suppliers.

\section{The One-sided Model}

\subsection{Model}
\label{ss:model}

Recall that in the one-sided model, we are given a bipartite graph $G$ with suppliers $S$ indexed by $i$ and
demanders $D$ indexed by $j$. Demander $j$ has a demand of $d_j$ that must be satisfied exactly, whereas 
supplier $i$ has single-peaked preferences with peak $s_i$; therefore, a 
supplier may be required to send more or less than his peak. In addition 
supplier $i$ is required to send at least $\ell_i$ and at most
$u_i$ units of flow; we may assume without loss of generality that 
$\ell_i \leq s_i \leq u_i$.  
The peaks of the demanders, their preferences, and the $\ell_i$ and $u_i$
are common knowledge; in contrast, for any supplier $i$, his peak $s_i$ and the set $f(i)$ of
demanders he is linked to may be private information held only by that supplier $i$ and hence must be
elicited by the mechanism.

Let $\lambda := (\lambda_i)_{i \in S}$ be non-negative.
Construct the following network $G(\lambda)$: introduce a source $s$ and
a sink $t$; arcs of the form $(s,i)$ for each supplier $i$ with capacity $\lambda_i$,
arcs of the form $(j,t)$ for each demander $j$ with capacity $d_j$; an infinite-capacity
arc from supplier $i$ to demander $j$ if supplier $i$ and demander $j$ share a link.
Let $\ell = (\ell_i)_{i \in S}$, $u = (u_i)_{i \in S}$, and $s = (s_i)_{i \in S}$. It is
straightforward to verify that the given problem admits a feasible solution if and only
if the maximum $s$-$t$ flow in $G(\ell)$ and $G(u)$ are, respectively, 
$\sum_{i \in S} \ell_i$ and $\sum_{j \in D} d_j$.
Consider now a maximum $s$-$t$ flow in the network $G(s)$.
By the max-flow min-cut theorem, there is a cut $C$ (a cut is a subset 
of nodes that contains the source $s$ but not the sink $t$) whose capacity 
is equal to that of the max-flow.
If the set of suppliers in $C$ is $X$ and the set of demanders in 
$C$ is $Y$, it is clear that $Y = f(X)$: if $Y \not \subseteq f(X)$, then
$C$ has infinite capacity, and if $Y \supset f(X)$ then $C$'s capacity can 
be improved by deleting the demanders in $Y \setminus f(X)$. Bochet et
al.~\cite{bim10} show that in any Pareto-optimal
allocation $x$ for the suppliers, $x_i \leq s_i$ for each $i \in X$ and
$x_i \geq s_i$ for each $i \in S \setminus X$.

If the min-cut is not unique, it is again well-known 
(see~\cite{lp88}) that there is a min-cut
with the largest $X$ (largest in the sense of inclusion), and a min-cut with the
smallest $X$ (again in the sense of inclusion). Call these 
sets $\overline{X}$ and $\underbar{X}$. It is easy to check that every 
supplier in $\overline{X} \setminus \underbar{X}$ will be at his
peak value in {\em all} Pareto optimal solutions. In the notation of 
Bochet et al.~\cite{bim10}, $M_{0} := \overline{X} \setminus \underbar{X}$, 
$M_{-} := \underbar{X}$, and $M_{+} := S \setminus \overline{X}$. To keep 
things simple, however, we shall dispense with
$M_0$ and use the partition $M_{-} = \underbar{X}$, 
$M_{+} = S \setminus \underbar{X}$.
In this case the partition of the demanders becomes 
$Q_{+} = f(M_{-})$ and $Q_{-} = D \setminus f(M_{-})$.
We note that our $M_{-}$ is still uniquely determined for each problem.
In what follows, often it will be important to talk about the 
set of suppliers involved in the cut, rather than the cut itself: we abuse
notation and talk about the cut $X$ when in fact the set of nodes in
the cut is really $s \cup X \cup f(X)$.

\subsection{Egalitarian Mechanism}
\label{ss:egal}

Suppose $(x_i)_{i \in S}$ is a Pareto optimal allocation. From the
earlier discussion it is clear that $x_i \in [s_i, u_i]$ for every supplier $i
\in M_+$,
and $x_i \in [\ell_i, s_i]$ for every supplier $i \in M_{-}$. 
Bochet et al.~\cite{bim10} prove that the egalitarian allocation,
which is defined independently
for the suppliers in $M_{-}$ and $M_{+}$,
Lorenz dominates all other Pareto optimal allocations.

For the suppliers in $M_{-}$, the egalitarian allocation is found by the
following
algorithm. Let $\lambda$ be a parameter whose value is increased continuously
from
zero, and let $m_i(\lambda) = {\rm median}(\ell_i, \lambda, s_i)$. Consider the
graph
$G(m(\lambda))$, where the capacity of the arc $(s,i)$ is $m_i(\lambda)$.
By the earlier discussion, we know that each supplier in $M_{-}$ will send at
least
$\ell_i$ and at most $s_i$ units of flow in a Pareto optimal solution, and that
every demander $j$ in $f(M_{-})$ will receive exactly $d_j$ units of flow.
We now study the sequence of networks $G(m(\lambda))$---specifically
the maximum $s$-$t$ flow in such networks---as $\lambda$ is increased from zero.
It is not hard to see that the maximum $s$-$t$ flow in $G(m(\lambda))$ 
is a weakly-increasing,
piecewise linear function of $\lambda$ with at most $2n$ breakpoints.
Moreover, each breakpoint is  one of the $\ell_i$, or one of the 
$s_i$ (type 1), or is associated with
a subset of suppliers $X$ such that 
\begin{equation}
\label{type2}
\sum_{i \in X} m_i(\lambda) \; = \; \sum_{j \in f(X)} d_j
\end{equation}
This we call a type-2 breakpoint.
At a type-1 breakpoint, the associated supplier is at his peak and so 
will not send any more flow (recall that every supplier $i \in M_{-}$ 
will send flow at most his peak $s_i$); at a type-2
breakpoint, however,
the group of suppliers in $X$ are sending enough flow to satisfy
the collective demand of the demanders in $f(X)$, so
 any further increase in flow from {\em any} supplier in $X$ would cause {\em
some}
demander in $f(X)$ to accept more than his peak demand.

If the given problem does not have any type-2 breakpoint, then the
egalitarian solution obtains by setting each supplier's allocation to his
peak value. Otherwise, let $\lambda ^{\ast }$ be the first type-2 breakpoint
of the max-flow function; by the max-flow min-cut theorem, for every subset $%
X$ satisfying (\ref{type2}) at $\lambda ^{\ast }$ the cut $C^{1}=\{s\}\cup X\cup
f(X)$ is a minimal cut in $G(m(\lambda ^{\ast }))$ providing a
certificate of optimality for the maximum-flow in $G (m(\lambda ^{\ast }))$%
. If there are several such cuts, we pick the one with the largest $X^{\ast
} $ (its existence is guaranteed by the usual supermodularity argument). The
egalitarian solution obtains by setting
\begin{equation*}
x_{i}={\rm median}(\ell_i, \lambda^{\ast }, s_{i}),\;\text{for}\;i\in X^{\ast
},\;\;\; y_{j}=d_{j},\;\text{for}\;j\in f(X^{\ast }),
\end{equation*}%
and assigning to other agents their egalitarian share in the reduced problem
involving the suppliers in $M_{-} \setminus X^{\ast}$ and the demanders in
$Q_{+} \setminus f(X^*)$. It is straightforward to verify that the first type-2
breakpoint $\lambda^{**}$ of this reduced problem will satisfy $\lambda^{**} >
\lambda^*$.

For the suppliers in $M_{+}$, a similar algorithm is used to 
determine the egalitarian allocation: here,
each demander $j \in Q_{-}$ receives exactly $d_i$ units of flow, 
whereas every supplier $i \in M_{+} = g(Q_{-})$ sends
at least $s_i$ and at most $u_i$ units of flow in a Pareto optimal solution. 
As before, we consider the graph
$G(m(\lambda))$, where the capacity of the arc $(s,i)$ is
$m_i(\lambda) := {\rm median}(s_i, \lambda, u_i)$. 
We increase $\lambda$ gradually and observe that the maximum
$s$-$t$ flow in $G(m(\lambda))$ is a weakly-increasing,
piecewise linear function of $\lambda$ with at most $2n$ breakpoints.
Moreover, each breakpoint is  one of the $s_i$, or one of 
the $u_i$ (type 1), or is associated with
a subset of suppliers $X$ such that 
\begin{equation*}
\sum_{i \in X} m_i(\lambda) \; = \; \sum_{j \in f(X)} d_j
\end{equation*}
This we call a type-2 breakpoint.
At a type-1 breakpoint, the associated supplier is at his 
upperbound and so cannot send any more
flow; at a type-2
breakpoint, however,
the group of suppliers in $X$ are sending enough just enough flow to satisfy
the collective demand of the demanders in $f(X)$, so
 any decrease in flow from {\em any} supplier in $X$ would cause {\em some}
demander in $f(X)$ to receive an amount strictly below his peak demand.
As before, if the given problem does not have any type-2 breakpoint, then the
egalitarian solution obtains by setting each supplier's allocation to his
upper bound. Otherwise, let $\lambda ^{\ast }$ be the first type-2 breakpoint
of the max-flow function, and let $X^*$ be the (largest) associated bottleneck
set of suppliers (as before). The
egalitarian solution obtains by setting
\begin{equation*}
x_{i}={\rm median}(s_i, \lambda^{\ast }, u_{i}),\;\text{for}\;i\in X^{\ast
},\;\;\; y_{j}=d_{j},\;\text{for}\;j\in f(X^{\ast }),
\end{equation*}%
and assigning to other agents their egalitarian share in the reduced problem
involving the suppliers in $M_{+} \setminus X^{\ast}$ and the demanders in
$Q_{-} \setminus f(X^*)$.
This completely defines the egalitarian solution.

\subsection{Group Strategyproofness}

We turn now to strategic aspects of the rationing problem with constraints. In
the one-sided
model, only the suppliers are modeled as ``agents,'' who possess potentially two
pieces of 
information that could be modeled as private: the
set of demanders they are compatible with, and their own preference over
allocations
\footnote{The set of demanders, their individual demands, as well as the 
the lower and upper bounds on arc-flows are assumed to be
common knowledge.}.
As the egalitarian mechanism is ``peak-only''~\cite{bim10}, it is sufficient
for the suppliers
to report only their peaks, rather than their entire preference ordering.

\begin{figure}[h]
\begin{center}
\includegraphics[width=100mm,height=50mm]{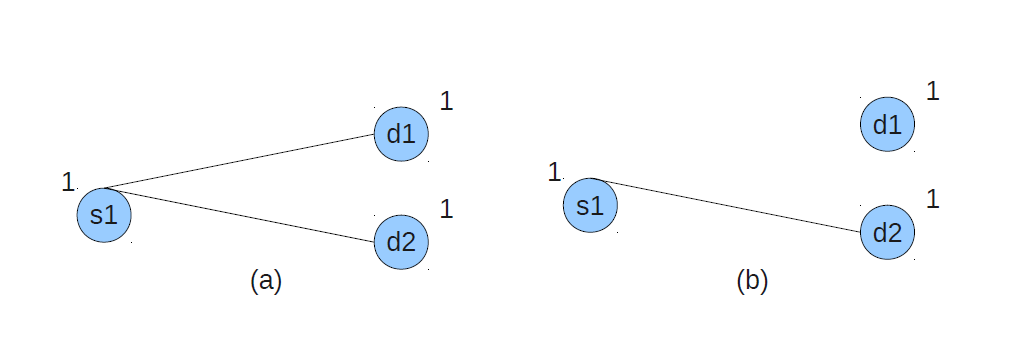}
\caption{Counterexample for Link SP}
\label{fig:lsp}
\end{center}
\end{figure}

It is a simple matter to verify that the egalitarian mechanism is {\em not} link
strategyproof. Consider
a supplier with a peak of 1, connected to two demanders, each with a demand of
1, see Figure~\ref{fig:lsp}.
If the supplier
reveals both links, his egalitarian allocation is 2, whereas by suppressing one
of the links, his 
egalitarian allocation improves to 1. Therefore in the rest of this section we
focus only on peak strategyproofness.

Bochet et al.~\cite{bim10} show that the egalitarian mechanism,
which finds the egalitarian allocation for any given
problem, is  peak strategyproof. Our main result in this section is that,
in fact, the egalitarian mechanism is peak groupstrategyproof. To set
the stage for this, we start with a lemma, which we first use to give
an alternative proof that the egalitarian mechanism is
strategyproof.

\begin{lemma}
\label{lem:structure2}
For a problem $(G,s,d)$, suppose the
decomposition is $M_{+}$ and $M_{-}$ (with $Q_{+}$, $Q_{-}$ defined
as before), and the
egalitarian allocation is
$x$. Consider the problem $(G,s',d)$ with $s'_j = s_j$ for all
$j \not = i$, with the decomposition being $M'_{+}$ and $M'_{-}$. 
\begin{itemize}
\item[(a)]
If $i \in M_{-}$ and $s'_i \geq s_i$, $M'_{+} = M_{+}$ and $M'_{-} = M_{-}$.
\item[(b)]
If $i \in M_{+}$ and $s'_i \leq s_i$, $M'_{+} = M_{+}$ and $M'_{-} = M_{-}$.
\end{itemize}
\end{lemma}

\noindent
{\bf Proof.}
By definition, $M_{-}$ is the smallest (both in terms of cardinality
and inclusion) min-cut in the graph $G(s)$ (see \S\ref{ss:model} 
for the definition). 
For $i \in M_{-}$, the arc $(s,i)$ does not 
contribute to the cut-capacity. If $s'_i \geq s_i$, the capacity of any cut
is weakly greater in $(G,s',d)$ than in $(G,s,d)$, whereas the capacity of
the cut $M_{-}$ stays the same, so part~(a) follows by the minimality
of $M_{-}$.  Similarly, for $i \in M_{+}$, the arc $(s,i)$ contributes to
cut-capacity, the capacity of the cut $M_{-}$ is smaller in $(G,s',d)$ than
in $(G,s,d)$ by {\em exactly} $s_i - s'_i$, whereas the capacity of any cut is weakly smaller in $(G,s',d)$
than in $(G,s,d)$ by {\em at most} $s_i - s'_i$. Again, part~(b) follows by the
minimality of $M_{-}$.
\qed

\begin{theorem}
\label{th:sp}
The egalitarian mechanism is peak strategyproof.
\end{theorem}

\noindent
{\bf Proof.}
For the problem $(G,s,d)$ let $x$ be the egalitarian allocation, and let
$M_+$ and $M_{-}$ be defined as before. Consider the problem $(G,s',d)$ with
$s_k = s'_k$ for all $k \not = i$. 
Suppose $i \in M_{-}$.
If $s'_i \geq s_i$, Lemma~\ref{lem:structure2} proves
that the decomposition does not change; it is easy to see that the egalitarian
allocation is unaffected as well, because the algorithm to compute operates
identically in the problems $(G,s,d)$ and $(G,s',d)$.
Similarly, if $i \in M_{+}$ and
$s'_i \leq s_i$, the decomposition does not change (by Lemma~\ref{lem:structure2}),
and the egalitarian allocation is unaffected as well.
Suppose agent $i$ reports $s'_i$ as his peak and the allocation changes
to $x'_i$. 
To prove strategyproofness, it suffices to show that any
$i \in M_{-}$ (weakly) prefers $x_i$ to $x'_i$ for all $s'_i < s_i$, and
that any $i \in M_{+}$ (weakly) prefers $x_i$ to $x'_i$ for all $s'_i >  s_i$.

Fix an $i \in M_{-}$, and suppose that $i$ reports a peak of $s'_i < s_i$.
In this case the decomposition {\em may} change; let $M'_{-}$ and 
$M'_{+}$ be the new decomposition. If $i \in M'_{-}$, an application of
Lemma~\ref{lem:structure2} to the problem $(G,s',d)$ shows that the
decomposition does not change, and that $x'_i = x_i$.
Suppose $i \in M'_{+}$. 
Let $D' := Q_{+} \cap Q'_{-}$, and $X' := M_{-} \cap M'_{+}$, and note
that by our supposition $X' \ni i$. Note also that 
$g(D') \cap M_{-} \subseteq X'$, as no agent in $M'_{-}$
has a link to any demander in $Q'_{-}$. Furthermore,
if $i \not \in g(D')$,
$x'_{i} = 0$, and again the result follows: recall that $f(i) \subseteq Q_{+}$;
and if $i \not \in g(D')$, $f(i) \subseteq Q'_{+}$, and the links from
$M'_{+}$ to $Q'_{+}$ do not carry any flow.
So we may assume that $i \in g(D')$.
We now make two simple observations
about the agents in $X' \cap g(D')$ in
the problem $(G,s',d)$: first every such agent
sends flow only to the demanders in $D'$, and therefore
$\sum_{k \in X' \cap g(D')} x'_k \; \leq \; \sum_{j \in D'} d_j$.
Also, as every agent in $X' \cap g(D')$ is (weakly) above his reported
peak, $x'_k \geq s_k$ for every $k \in X' \cap g(D')$, $k \not = i$,
and $x'_i \geq s'_i$.
This implies
\begin{equation}
\label{eq:lb}
\sum_{k \in X' \cap g(D'), k \not = i} s_k + x'_i \leq \sum_{j \in D'} d_j.
\end{equation}
We next claim that in the problem $(G,s,d)$,
$\sum_{k \in X' \cap g(D')} x_k = \sum_{j \in D'} d_j$. To see why,
observe that
the demands of $D'$ are covered in the problem $(G,s,d)$ by the
suppliers in $M_{-} \cap g(D')$; but every demander in $D'$ moves
from $Q_{+}$ to $Q'_{-}$, so every supplier in $M_{-} \cap g(D')$
must move to $M'_{+}$ (as there cannot be an
edge between a supplier in $M'_{-}$ and a demander in $Q'_{-}$).
This implies that any supplier supplying a positive amount to
a demander in $D'$ in the problem $(G,s,d)$ must be in $X' \cap g(D')$.
Note also
that for each $k \in X' \cap g(D')$, $x_k \leq s_k$. 
These, along with $X' \cap g(D') \subseteq M_{-}$, imply
\begin{equation}
\label{eq:ub}
\sum_{k \in X' \cap g(D'),  k \not = i} s_k + x_i \geq \sum_{j \in D'} d_j.
\end{equation}
Inequalities (\ref{eq:lb}) and (\ref{eq:ub}) imply $x'_i \leq x_i$, as required.

Now fix an $i \in M_{+}$, and suppose that $i$ reports a peak of $s'_i < s_i$.
In this case the decomposition {\em may} change; let $M'_{-}$ and 
$M'_{+}$ be the new decomposition. If $i \in M'_{+}$, as before, an application of
Lemma~\ref{lem:structure2} to the problem $(G,s',d)$ shows that
the decomposition does not change, and that $x'_i = x_i$.
Suppose $i \in M'_{-}$. 
Let $D' := Q_{-} \cap Q'_{+}$, and $X' := M_{+} \cap M'_{-}$, and note
that by our supposition $X' \ni i$. Note also that 
$f(X') \cap Q_{-} \subseteq D'$, as no agent in $M'_{-}$ can have
a link to any demander in $Q'_{+}$.
We now make two simple observations
about the demanders in $f(X') \cap D'$ in the problem $(G,s',d)$: first every such demander
can receive flow only from the agents in $X'$, and therefore
$\sum_{k \in X'} x'_k \; \geq \; \sum_{j \in f(X') \cap D'} d_j$.
Also, as every agent in $X'$ is (weakly) below his reported
peak (in the new problem), $x'_k \leq s_k$ for every $k \in X'$, $k \not = i$,
and $x'_i \leq s'_i$.
This implies
\begin{equation}
\label{eq:lbb}
\sum_{k \in X',  k \not = i} s_k + x'_i \; \geq \; \sum_{j \in f(X') \cap D'} d_j.
\end{equation}
We next claim that in the problem $(G,s,d)$,
$\sum_{k \in X'} x_k = \sum_{j \in f(X') \cap D'} d_j$: in $(G,s,d)$ the
suppliers in $X'$ send flow only to the demanders in $f(X') \cap D'$, who
receive flow only from these suppliers. Furthermore,
$x_k \geq s_k$ for each $k \in X'$. In particular,
\begin{equation}
\label{eq:ubb}
\sum_{k \in X', k \not = i} s_k + x_i \leq \sum_{j \in f(X') \cap D'} d_j.
\end{equation}
Inequalities (\ref{eq:lbb}) and (\ref{eq:ubb}) imply $x'_i \geq x_i$, as required.
\qed

In fact, the ideas in the proof of Theorem~\ref{th:sp} can be used to prove the
following result, which weakens the conditions under which the decomposition
is guaranteed not to change.

\begin{lemma}
\label{lem:structure}
For a problem $(G,s,d)$, suppose the
decomposition is $M_{+}$ and $M_{-}$ (with $Q_{+}$, $Q_{-}$ defined
as before), and the
egalitarian allocation is
$x$. Consider the problem $(G,s',d)$ with $s'_j = s_j$ for all
$j \not = i$, with the decomposition being $M'_{+}$ and $M'_{-}$. 
\begin{itemize}
\item[(a)]
If $i \in M_{-}$ and $s'_i > x_i$, $M'_{+} = M_{+}$ and $M'_{-} = M_{-}$.
\item[(b)]
If $i \in M_{+}$ and $s'_i < x_i$, $M'_{+} = M_{+}$ and $M'_{-} = M_{-}$.
\end{itemize}
\end{lemma}

\noindent
{\bf Proof.}
By definition, $M_{-}$ is the smallest (both in terms of cardinality
and inclusion) min-cut in the graph $G(s)$ (see \S\ref{ss:model} 
for the definition). 
For $i \in M_{-}$, the arc $(s,i)$ does not 
contribute to the cut-capacity. If $s'_i \geq s_i$, the capacity of any cut
is weakly greater in $(G,s',d)$ than in $(G,s,d)$, whereas the capacity of
the cut $M_{-}$ stays the same, so the result
follows. Suppose now that $x_i < s'_i < s_i$, the max $s$-$t$ flow in $G(s')$ is
weakly below that of $G(s)$, but the egalitarian allocation $x$ is still
feasible, so $x$ continues to be a max-flow, so $M_{-}$ continues to be a
min-cut in $G(s')$. We need to show that it remains the minimal min-cut.
First observe that $M'_{-} \subseteq M_{-}$,
as $M'_{-}$ is the minimal min-cut in $G(s')$ whereas $M_{-}$ is {\em a} 
min-cut for $G(s')$. If $i \in M'_{-}$, then the capacity of the cut
$M'_{-}$ is the same in $G(s)$ and $G(s')$, so the minimality of $M_{-}$
in the problem $(G,s,d)$ implies $M'_{-} = M_{-}$. Suppose $i \not \in M'_{-}$.
Let $X = M_{-} \setminus M'_{-}$, and note that $i \in X$. Note also that
$Q_{+} = f(M_{-})$ and $Q'_{+} = f(M'_{-})$, so that the net change in
the cut capacity when the suppliers in $X$ move from $M_{-}$ to $M'_{+}$
is precisely
$\sum_{k \in X} s'_{k} - \sum_{j \in Q_{+} \setminus Q'_{+}} d_j$.
In the problem $(G,s,d)$, however, the demanders in $Q_{+}  \setminus Q'_{+}$
receive flow
only from the suppliers in $X$, each of whom sends no more than his
peak: thus, $\sum_{k \in X} x_{k} \geq \sum_{j \in Q_{+} \setminus Q'_{+}} d_j$,
and
$s_k \geq x_k$ for each $k$. An easy implication is that
$s'_k \geq x_k$ for each $k \in X$, $k \neq i$, and $s'_i > x_i$.
Thus the net change in cut capacity in moving from $M_{-}$ to $M'_{-}$ is
strictly positive, which
implies $M'_{-}$ cannot be a min-cut.
A similar argument establishes part~(b).
\qed

We conclude this section with a proof  that the egalitarian mechanism is, in fact, group strategyproof.

\begin{theorem}
The egalitarian mechanism is peak groupstrategyproof.
\end{theorem}

\noindent
{\bf Proof.} 
Suppose not. Focus on a counterexample $G$ with the
{\em smallest} number of nodes. Suppose the true peaks
of the suppliers are $s$ and suppose they misreport
their peaks to be $s'$. Fix a coalition $A$ of
agents: note that this
coalition includes all the agents $k$ with $s'_k \not = s_k$.
Let $x$ and $x'$ be the respective allocations
to the agents when they report $s$ and $s'$ respectively. As with
the earlier proof, let $M_{+}, M_{-}$ be the decompsition 
when the agents report $s$, and let $M'_{+}, M'_{-}$ 
be the decomposition when the agents report $s'$.
We shall show that when the
agents report $s'$ rather than $s$
the only allocation in which each agent in $A$
is (weakly) better off is one in which $x'_k = x_k$ for all $k \in A$,
establishing the required contradiction.

Let $D' := Q_{+} \cap Q'_{-}$.
Note that $g(D') \subseteq M'_{+}$, for otherwise there will be a supplier
in $M'_{-}$ with a link to a demander in $Q'_{-}$.
We now make two simple observations about the agents in $M_{-} \cap g(D')$: 
\begin{itemize}
\item
When the report is $s'$, every such agent can send flow only to the demanders 
in $D'$: this is because $f(M_{-}) \subseteq Q_{+}$, and each agent in
$g(D')$ can send flow only to the agents in $Q'_{-}$. Therefore
$\sum_{k \in M_{-} \cap g(D')} x'_k \; \leq \; \sum_{j \in D'} d_j$.
\item
When the report is $s$, the demanders in $D'$ can receive flow only
from such agents: the demanders in $D'$ can receive flow only from the
suppliers in $M_{-}$ and they are connected only to the suppliers in
$g(D')$. Therefore
$\sum_{k \in M_{-} \cap g(D')} x_k \geq \sum_{j \in D'} d_j$.
\end{itemize}
Note also that $s'_k \leq x'_k$ and $x_k \leq s_k$ for any $k \in M_{-} \cap g(D')$,
and that $s'_k = s_k$ for all $k \not \in A$.
These observations lead to
\begin{equation}
\label{eq:lbg}
\sum_{\substack{ k \in M_{-} \cap g(D') \\ k \not \in A} } s_k + \sum_{\substack{k \in M_{-} \cap g(D') \\ k \in A}} x'_k  \; = \;
   \sum_{\substack{ k \in M_{-} \cap g(D') \\ k \not \in A} } s'_k + \sum_{\substack{k \in M_{-} \cap g(D') \\ k \in A}} x'_k  \; \leq \;
    \sum_{k \in M_{-} \cap g(D')} x'_k \; \leq \; \sum_{j \in D'} d_j,
\end{equation}
and
\begin{equation}
\label{eq:ubg}
\sum_{j \in D'} d_j \; \leq \; \sum_{k \in M_{-} \cap g(D')} x_k \; \leq \; 
\sum_{\substack{ k \in M_{-} \cap g(D') \\ k \not \in A} } s_k + \sum_{\substack{k \in M_{-} \cap g(D') \\ k \in A}} x_k.
\end{equation}

\bigskip

For every agent in $A$ to be (weakly) better off when reporting $s'$, we must have
$x'_k \geq x_k$ for each $k \in A$. Combining this with
inequalities (\ref{eq:lbg}) and (\ref{eq:ubg}), we conclude that 
$x'_k = x_k$ for each $k \in M_{-} \cap g(D') \cap A$. Moreover, these inequalities
also imply that $x'_k = x_k = s_k$ for each $k \in M_{-} \cap g(D')$, $k \not \in A$.
Thus, $x'_k = x_k$ for all $k \in M_{-} \cap g(D')$. 
Also, whether the report is $s$ or is $s'$, the suppliers in 
$M_{-} \cap g(D')$ send all of their
flow only to the demanders in $D'$; and that these demanders
receive all of their flow only from the suppliers in $M_{-} \cap g(D')$.
Therefore, removing the suppliers in $M_{-} \cap g(D')$ and the 
demanders in $D'$
does not affect the egalitarian solution for either problem. As we picked a
smallest counterexample, $D' = \emptyset$.

We now turn to the other case.
Let $\tilde{X} := M_{+} \cap M'_{-}$.
Note that
$f(\tilde{X}) \cap Q_{-} \subseteq Q'_{+}$, for otherwise there will be
a supplier in $M'_{-}$ linked to a demander in $Q'_{-}$.
Consider the demanders in $f(\tilde{X}) \cap Q_{-}$: 
\begin{itemize}
\item
When the report is $s'$, every such demander can receive flow only from
the suppliers in $\tilde{X}$: such demanders are linked only to the
suppliers in $M_{+}$ and can receive flow only from the suppliers in
$M'_{-}$. Therefore
$\sum_{k \in \tilde{X}} x'_k \; \geq \; \sum_{j \in f(\tilde{X}) \cap Q_{-}} d_j$.
\item
When the report is $s$, the suppliers in $\tilde{X}$ send flow
only to the demanders in $Q_{-}$, and they can send flow only to the
demanders they are connected to, so the suppliers in $\tilde{X}$ can
send flow only to the demanders in $f(\tilde{X}) \cap Q_{-}$. Therefore
$\sum_{k \in \tilde{X}} x_k \leq \sum_{j \in f(\tilde{X}) \cap Q_{-}} d_j$.
\end{itemize}
Note also that
$s'_k \geq x'_k$ and $x_k \geq s_k$ for any $k \in \tilde{X}$,
and that $s'_k = s_k$ for all $k \not \in A$.
Putting all this together we have:
\begin{equation}
\label{eq:lbbg}
\sum_{k \in \tilde{X} \setminus A} s_k + \sum_{k \in \tilde{X} \cap A} x'_k \; = \; 
\sum_{k \in \tilde{X} \setminus A} s'_k + \sum_{k \in \tilde{X} \cap A} x'_k \; \ge \; 
\sum_{k \in \tilde{X}} x'_k  \; \geq \;
\sum_{j \in f(\tilde{X}) \cap Q_{-}} d_j,
\end{equation}
and
\begin{equation}
\label{eq:ubbg}
\sum_{j \in f(\tilde{X}) \cap Q_{-}} d_j \; \geq 
	\; \sum_{k \in \tilde{X}} x_k \; \geq \;
\sum_{k \in \tilde{X} \setminus A} s_k + \sum_{i \in \tilde{X} \cap A} x_i.
\end{equation}

\bigskip

For every agent in $A$ to be (weakly) better off when reporting $s'$, we must have
$x_k \leq x'_k$ for each $k \in A$. Combining this with
inequalities (\ref{eq:lbbg}) and (\ref{eq:ubbg}), we conclude that 
$x'_k = x_k$ for each $k \in \tilde{X} \cap A$. Moreover, these inequalities
also imply that $x'_k = x_k = s_k$ for each $k \in \tilde{X} \setminus A$.
Thus, $x'_k = x_k$ for all $k \in \tilde{X}$. 
Note that the suppliers in $\tilde{X}$ send all of their flow
flow to the demanders in $f(\tilde{X}) \cap Q_{-}$, whether the report is $s$ 
or $s'$; also the demanders in $f(\tilde{X}) \cap Q_{-}$ receive all of
their flow from the suppliers in $\tilde{X}$, whether the report is $s$ or
$s'$. Therefore,
removing the suppliers in $\tilde{X}$ and the demanders in 
$f(\tilde{X}) \cap Q_{-}$
does not affect the egalitarian solution for either problem. As we picked a
smallest counterexample, $\tilde{X} = \emptyset$.

We now establish that the decomposition does not change in a smallest
counterexample. We already know that $D' = \emptyset$, which implies
$Q'_{-} \subseteq Q_{-}$. Suppose this containment is strict so that
there is a demander $j \in Q_{-} \setminus Q'_{-}$. Then, 
$g(j) \subseteq M_{+}$. As $\tilde{X} = \emptyset$, $g(j) \subseteq M'_{+}$,
which implies demander $j$ cannot receive any flow when the report is $s'$.
Therefore $Q'_{-} = Q_{-}$, which implies $Q'_{+} = Q_{+}$,
$M'_{+} = M_{+}$, and $M'_{-} = M_{-}$.

To complete the argument, let $A$ be as defined earlier.
Let $A_{+} = A \cap M_{+}$ and $A_{-} = A \cap M_{-}$.
For any $i \in A_{-}$, $s'_i < x_i$ implies $x'_i \leq s'_i < x_i$, causing
$i$ to do worse by reporting $s'_i$.
Likewise, any $i \in A_{+}$, $s'_i > x_i$ implies $x'_i \geq s'_i > x_i$, causing
$i$ to do worse by reporting $s'_i$.
So any improving coalition $A$ must be such
that $s'_i \geq x_i$ for all $i \in A_{-}$ and
$s'_i \leq x_i$ for all $i \in A_{+}$.
But in this case the egalitarian solution does not change for either problem.
\qed

\section{The Two-sided Model}

\subsection{Model}
\label{ss:model2}

We now turn to the two-sided model, introduced by Bochet et
al.~\cite{bims10}.
This is closely related to the one-sided model, and yet there are some
important differences, so our treatment will be concise, and will focus
mostly on the aspects that make this model different. As in the one-sided
model, we are given a bipartite graph $G$ with suppliers ($M$) indexed by $i$
and demanders by $j$. In the two-sided model, however, both the suppliers
and demanders have single-peaked preferences: supplier $i$ has a peak
of $s_i$,  and demander $j$ has a peak of $d_j$.  Define the graph
$\Gamma(G,s,d)$ by adding a source 
$\sigma $ connected to all suppliers,
and a sink $\tau $ connected to all demanders; by orienting the edges from
source to sink; by setting the capacity of an edge in $G$ to infinity, that
of an edge $\sigma i,i\in S$, to $s_{i}$, and that of $j\tau ,\tau \in D$,
to $d_{j}$. 
A $\sigma $-$\tau $ cut (or simply a cut) in this graph is a
subset $C$ of nodes that contains $\sigma $ but not $\tau $. The capacity of
a cut $C$ is the total capacity of the edges that are oriented from a node
in $C$ to a node outside of $C$ (such edges are said to be \textquotedblleft
in the cut\textquotedblright ).
It is well-known that the max-flow from $\sigma$ to $\tau$ in this network is
the same as the capacity of a min $\sigma$-$\tau$ cut. Fix an arbitrary
min $\sigma$-$\tau$ cut $C$, and let $X$ be the set of suppliers in $C$
and $Y$ be the set of demanders in $C$. It is not hard to see that $Y = f(X)$
in a min-cut. Bochet et al.~\cite{bims10} characterize the Pareto optimal
solutions to this model using a network construction very similar to that used
for the one-sided model. In particular they show that in any Pareto optimal
allocation $(x,y)$, $x_i \leq s_i$ for $i \in X$ and $x_i \geq s_i$ for each
supplier $i \not \in X$; similarly, $y_j \leq d_j$ for all $j \in Y$, and
$y_j \geq d_j$ for all $j \not \in Y$. 
In addition they show that in a Pareto-optimal allocation, there is
no flow from a supplier {\em not} in $X$ to a demander in $Y$ (even though links
may exist between such suppliers and demanders). In effect, the problem
decomposes into two subproblems, one involving the suppliers in $X$ and the
demanders in $Y$;the other involving the suppliers {\em not} in $X$ and the
demanders {\em not} in $Y$.  
The egalitarian allocation is found
independently for each subproblem, using exactly the same algorithm described
earlier in \S\ref{ss:egal}. 
They also show that the egalitarian allocation can be found by an alternative
algorithm as follows: first, set all the demands to be $d_j$ and apply the
one-sided egalitarian algorithm described earlier in \S\ref{ss:egal} with the
additional restriction that no supplier sends more than his peak, 
and that no demander receives more than hers; this gives the egalitarian
allocation for the suppliers. The egalitarian allocation for the demanders
is obtained by applying the same algorithm under the same additional 
restriction, but interchanging the roles of the suppliers and demanders 
(thus, we fix the supplies to be the $s_i$ etc.) It is this equivalent
definition that we will use in the proofs.
A key result is that the egalitarian allocation is
the Lorenz dominant element of the subset of the Pareto optimal allocations in
which no agent sends or receives more than their peak (this set is called
Pareto$^*$ by Bochet et al.). By picking a cut $C$ with the largest $X$,
we find a canonical decomposition for each problem, as such a cut is 
unique: to relate this with the notation of Bochet et al., the $X$ 
corresponding to the largest $C$ is $S_{-}$, and the demanders $Y$ in
$C$ is the set $D_{+}$; the suppliers outside of $X$ belong to $S_{+}$,
and the demanders outside of $Y$ are in $D_{-}$.

We turn now to strategic issues. Bochet et al.~\cite{bims10} show
that the egalitarian mechanism is both link strategyproof and peak
strategyproof. Here we show that the egalitarian mechanism is in fact
peak groupstrategyproof.

\begin{theorem}
In the two-sided model, the egalitarian mechanism is peak group strategyproof.
\end{theorem}

\noindent
{\bf Proof.}
Suppose not. Focus on a counterexample $G$ with the
{\em smallest} number of nodes. Suppose the true peaks
of the suppliers and demanders are $s$ and $d$ respectively,
and suppose their respective misreports are $s'$ and $d'$.
We can assume that $d_j > 0$ for every 
demander $j$, as otherwise deleting $j$ would result in a smaller counterexample.
Fix a coalition $A$ of suppliers and a coalition $B$ of
demanders : note that $A$ contains
all the suppliers $k$ with $s'_k \not = s_k$, and $B$
includes all demanders $\ell$ with $d'_\ell \not = d_\ell$.

Let $(x,y)$ and $(x',y')$ be the respective allocations
to the suppliers and demanders when they report $(s,d)$ and $(s',d')$ 
respectively. 
Let $S_{+}, S_{-}, D_{+}, D_{-}$ be the decompsition 
when the agents report $(s,d)$, and let $S'_{+}, S'_{-},
D'_{+}, D'_{-}$ be the decomposition when the agents 
report $(s',d')$.
We shall show that when
the agents report $(s',d')$ rather than $(s,d)$,
the only allocation in which each agent in $A \cup B$
is (weakly) better off, then $x'_k = x_k$ for 
all $k \in A$ and $y'_{\ell} = y_{\ell}$ for all $\ell \in B$.
This establishes the required contradiction.

Let $Y' := D_{+} \cap D'_{-}$.
Note that $g(Y') \subseteq S'_{+}$, for, otherwise, there will
be a supplier in $S'_{-}$ with a link to a demander in $D'_{-}$.
We now make two simple observations about the suppliers
in $S_{-} \cap g(Y')$: 
\begin{itemize}
\item
For any such supplier $k$, $s'_k = x'_k$ and $x_k \leq s_k$. Also,
$d_\ell = y_\ell$ and $y'_\ell \leq d'_\ell$ for any $\ell \in Y'$.
\item
When the report is $s'$, every such supplier can send flow only to the demanders 
in $Y'$: this is because $f(S_{-}) \subseteq D_{+}$, and each supplier in
$g(Y')$ can send flow only to the agents in $D'_{-}$. Therefore
$\sum_{k \in S_{-} \cap g(Y')} x'_k \; \leq \; \sum_{\ell \in Y'} y'_\ell$.
\item
When the report is $s$, the demanders in $Y'$ can receive flow only
from such suppliers: the demanders in $Y'$ can receive flow only from the
suppliers in $S_{-}$ and they are connected only to the suppliers in
$g(Y')$. Therefore
$\sum_{k \in S_{-} \cap g(Y')} x_k \geq \sum_{\ell \in Y'} y_\ell$.
\end{itemize}
Finally, note that $s'_k = s_k$ for all $k \not \in A$, and
$d'_\ell = d_\ell$ for all $\ell \not \in B$.
These observations first lead to
\begin{equation}
\label{eq:2glbg}
\sum_{\substack{ k \in S_{-} \cap g(Y') \\ k \not \in A} } s_k + 
\sum_{\substack{k \in S_{-} \cap g(Y') \\ k \in A}} x'_k  \; = \;
   \sum_{\substack{ k \in S_{-} \cap g(Y') \\ k \not \in A} } s'_k +
\sum_{\substack{k \in S_{-} \cap g(Y') \\ k \in A}} x'_k  \; = \;
    \sum_{k \in S_{-} \cap g(Y')} x'_k \; \leq \; \sum_{\ell \in Y'} y'_\ell.
\end{equation}
Note that every demander $\ell$ in $Y' \cap B$ receives {\em exactly} his peak
allocation $d_\ell$ for a truthful report, so for the coalition $B$ of 
demanders to do weakly better in the $(G,s',d')$ problem, 
$y'_\ell = d_\ell$ for each such $\ell$.
Therefore,
\begin{equation}
\label{eq:2gs}
\sum_{\ell \in Y'} y'_\ell \; = \; 
     \sum_{\ell \in Y' \setminus B} y'_\ell + \sum_{\ell \in Y' \cap B} y'_\ell \; \leq \; 
     \sum_{\ell \in Y' \setminus B} d'_\ell + \sum_{\ell \in Y' \cap B} d_\ell \; = \; 
     \sum_{\ell \in Y'} d_\ell.
\end{equation}
Finally,
\begin{equation}
\label{eq:2gubg}
\sum_{\ell \in Y'} d_\ell \; = \; \sum_{\ell \in Y'} y_{\ell} \; \leq \;
\sum_{k \in S_{-} \cap g(Y')} x_k \; \leq \; 
\sum_{\substack{ k \in S_{-} \cap g(Y') \\ k \not \in A} } s_k + 
\sum_{\substack{k \in S_{-} \cap g(Y') \\ k \in A}} x_k.
\end{equation}

\bigskip

For every supplier in $A$ to be (weakly) better off when 
reporting $s'$, we must have
$x'_k \geq x_k$ for each $k \in S_{-} \cap g(Y')$. Combining this with
inequalities (\ref{eq:2glbg}) and (\ref{eq:2gubg}), we conclude that 
all the inequalities in (\ref{eq:2glbg})-(\ref{eq:2gubg}) hold as
equations.
In particular, $x'_k = x_k$ for all $k \in S_{-} \cap g(D')$, and
$y'_{\ell} = y_{\ell}$ for $\ell \in Y'$.
Therefore, whether the report is $s$ or is $s'$, the suppliers in 
$S_{-} \cap g(Y')$ send all of their
flow only to the demanders in $Y'$; and that these demanders
receive all of their flow only from the suppliers in $S_{-} \cap g(Y')$.
Therefore, removing the suppliers in $S_{-} \cap g(Y')$ and the 
demanders in $Y'$
does not affect the egalitarian solution for either problem. As we picked a
smallest counterexample, $Y'$ must be empty.

We now turn to the other case.
Let $\tilde{X} := S_{+} \cap S'_{-}$.
Note that
$f(\tilde{X}) \cap D_{-} \subseteq D'_{+}$, for otherwise there will be
a supplier in $S'_{-}$ linked to a demander in $D'_{-}$.
Consider the demanders in $f(\tilde{X}) \cap D_{-}$: 
\begin{itemize}
\item
For any such demander $\ell$, $d'_\ell = y'_\ell$ and $y_\ell \leq d_\ell$. 
Also, $s_k = x_k$ and $x'_k \leq s'_k$ for any $k \in \tilde{X}$.
\item
When the report is $s'$, every such demander can receive flow only from
the suppliers in $\tilde{X}$: such demanders are linked only to the
suppliers in $S_{+}$ and can receive flow only from the suppliers in
$S'_{-}$. Therefore
$\sum_{k \in \tilde{X}} x'_k \; \geq \; 
	\sum_{\ell \in f(\tilde{X}) \cap D_{-}} y'_\ell$.
\item
When the report is $s$, the suppliers in $\tilde{X}$ send flow
only to the demanders in $D_{-}$, and they can send flow only to the
demanders they are connected to, so the suppliers in $\tilde{X}$ can
send flow only to the demanders in $f(\tilde{X}) \cap D_{-}$. Therefore
$\sum_{k \in \tilde{X}} x_k \leq 
	\sum_{\ell \in f(\tilde{X}) \cap D_{-}} y_\ell$.
\end{itemize}
Finally, note that $s'_k = s_k$ for all $k \not \in A$, and
$d'_\ell = d_\ell$ for all $\ell \not \in B$.
Putting all this together, we have:
\begin{equation}
\label{eq:2glbbg}
\sum_{\substack{ \ell \in f(\tilde{X}) \cap D_{-}) 
\\ \ell \not \in B} } d_\ell + 
\sum_{\substack{ \ell \in f(\tilde{X}) \cap D_{-}) 
\\ \ell \in B} } d'_\ell 
 \; = \;
\sum_{\ell \in f(\tilde{X}) \cap D_{-}} d'_\ell \; = \;
\sum_{\ell \in f(\tilde{X}) \cap D_{-}} y'_\ell,
\end{equation}
and
\begin{equation}
\label{eq:2gaux}
\sum_{\ell \in f(\tilde{X}) \cap D_{-}} y'_\ell \; \leq \;
\sum_{k \in \tilde{X}} x'_k  \; \leq \;
\sum_{k \in \tilde{X} \setminus A} s'_k + 
	\sum_{k \in \tilde{X} \cap A} x'_k \; = \; 
\sum_{k \in \tilde{X} \setminus A} s_k + 
\sum_{k \in \tilde{X} \cap A} x'_k.
\end{equation}
Note that every supplier $k$ in $\tilde{X} \cap A$ receives {
\em exactly} his peak
allocation $s_k$ for a truthful report, so for the coalition $A$ of 
suppliers to do weakly better in the $(G,s',d')$ problem, 
$x'_k = s_k$ for each such $k$.
Thus,
\begin{equation}
\label{eq:2gubbg}
\sum_{k \in \tilde{X} \setminus A} s_k + 
\sum_{k \in \tilde{X} \cap A} x'_k \; = \;
\sum_{k \in \tilde{X}} s_k  \; = \;
\sum_{k \in \tilde{X}} x_k \; \leq \;
\sum_{\ell \in f(\tilde{X}) \cap D_{-}} y_\ell \; \leq \;
\sum_{\substack{ \ell \in f(\tilde{X}) \cap D_{-}
\\ \ell \not \in B} } d_\ell + 
\sum_{\substack{ \ell \in f(\tilde{X}) \cap D_{-}
\\ \ell \in B} } y_\ell 
\end{equation}

\bigskip
For every demander in $B$ to be (weakly) better off, we must have
$y'_{\ell} \geq y_{\ell}$ for each $\ell \in f(\tilde{X}) \cap D_{-}$.
Combining this with
inequalities (\ref{eq:2glbbg})-(\ref{eq:2gubbg}), we conclude that 
all the inequalities in (\ref{eq:2glbbg})-(\ref{eq:2gubbg}) hold as
equations.
In particular, $x'_k = x_k$ for all $k \in \tilde{X}$, and
$y'_{\ell} = y_{\ell}$ for $\ell \in f(\tilde{X}) \cap D_{-}$.
Therefore, whether the report is $s$ or is $s'$, the suppliers in 
$\tilde{X}$ send all of their
flow only to the demanders in $f(\tilde{X}) \cap D_{-}$; and that these demanders
receive all of their flow only from the suppliers in $\tilde{X}$.
Therefore, removing the suppliers in $\tilde{X}$ and the 
demanders in $f(\tilde{X}) \cap D_{-}$
does not affect the egalitarian solution for either problem. As 
we picked a smallest counterexample, $\tilde{X}$ must be empty.

We now establish that the decomposition does not change in a smallest
counterexample. We already know that $Y' = \emptyset$, which implies
$D'_{-} \subseteq D_{-}$. Suppose this containment is strict so that
there is a demander $j \in D_{-} \setminus D'_{-}$. Then, 
$g(j) \subseteq S_{+}$. As $\tilde{X} = \emptyset$, $g(j) \subseteq S'_{+}$,
which implies demander $j$ cannot receive any flow when the report is $s'$
(i.e. $x'_{j} = 0$). This is a contradiction since, $d'(j) > 0$, then the
egalitarian solution allocates the pareto value $x'_{j} = d'_{j}$ for all $j \in
D'_{+}$. (w.l.o.g we can skip the case $d'_{j} = 0$ as we can delete such a $j$
 to obtain the new decomposition or just place it in $D_{-}$). Therefore $D'_{-}
= D_{-}$, which implies $D'_{+} = D_{+}$, $S'_{+} = S_{+}$, and $S'_{-} =
S_{-}$.

To complete the argument, let $A$ be as defined earlier.
Let $A_{+} = A \cap S_{+}$ and $A_{-} = A \cap S_{-}$, $B_{+} = A \cap D_{+}$
and $B_{-} = A \cap D_{-}$. Now, for any $j \in B_{+}$, $d'_{j} \neq d_{j}$
implies $y'_{j} = d'_{j} \neq d_{j}$ causing $j$ to do worse by reporting
$d'_{j}$. Hence, it follows, $\forall j \in B_{+}$, $d'_{j} = d_{j}$. By a
similar argument, we could establish $s'_{j} = s_{j} \forall j \in A_{+}$.
 
For any $i \in A_{-}$, $s'_i < x_i$ implies $x'_i \leq s'_i < x_i$, causing
$i$ to do worse by reporting $s'_i$.
Likewise, any $i \in B_{-}$, $d'_i < y_i$ implies $y'_i \leq d'_i < y_i$,
causing $i$ to do worse by reporting $d'_i$. So any improving coalition $A$ must
be such that $s'_i \geq x_i$ for all $i \in A_{-}$ and
$d'_i \geq y_i$ for all $i \in B_{-}$.
But in this case the egalitarian solution does not change for either problem.
\qed

\begin{figure}[h]
\begin{center}
\includegraphics[width=100mm,height=50mm]{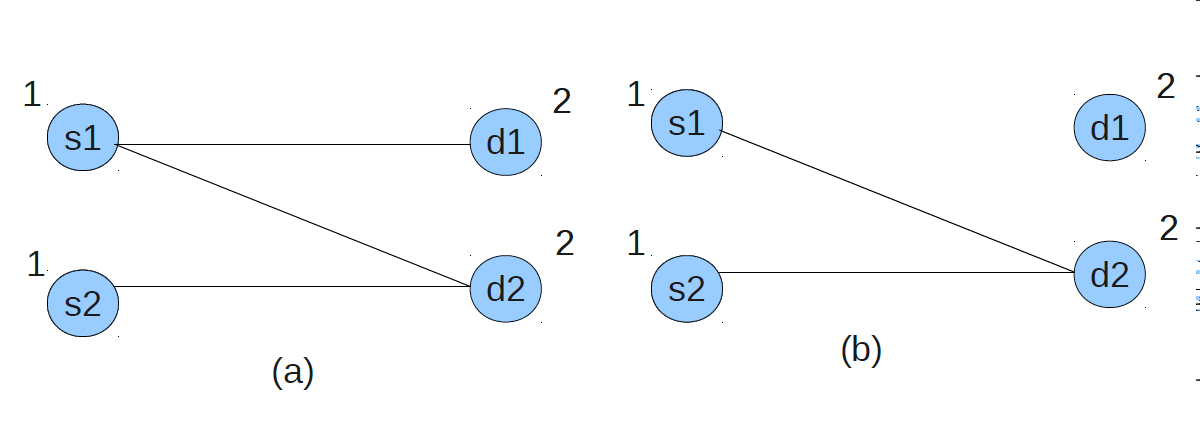}
\label{fig:2lsp}
\caption{A counterexample for link GSP}
\end{center}
\end{figure}

We turn now to link group strategyproofness. That the egalitarian mechanism is
{\em not} link group strategyproof in the two-sided model is not difficult
to see. Consider the network shown in Figure 2. The network~(a)
represents the true network, with the peaks shown next to the agent labels.
The egalitarian allocation gives 1 unit to each supplier and to each demander
on this example. Suppose however supplier 1 and demander 2 collude, and supplier
1 does not report his link to demander 1. In the resulting network, shown in~(b),
each supplier still receives his peak allocation; demander 2 now receives
her peak, and demander 1 receives nothing. Note that both members of the coalition
weakly improve, and demander 2 strictly improves, proving that the egalitarian
mechanism is in general {\em not} link group strategyproof
\footnote{This example may suggest that if we require each member of the
deviating coalition to strictly improve their allocation, then the egalitarian
mechanism may be link group strategyproof. However, this is also false, as shown
by Bogomolnaia and Moulin~\cite{bm04}. They construct an example involving 4
agents on each side with all peaks identically 1 in which a coalition of 
agents from both sides deviate and all {\em strictly} improve.}. 
The following result,
however, shows that the egalitarian mechanism satisfies a limited form of link
group strategyproofness.

\begin{theorem}
In the two sided model, the egalitarian mechanism is link group strategy proof
when the coalition is restricted to the set of suppliers only (demanders only).
\end{theorem}

\noindent
{\bf Proof.} We prove the result for an arbitrary coalition of suppliers; the
result
for the demanders follow by a similar argument. Let $A_i$ be the set of
demanders
that supplier $i$ is linked to, and let $A'_i$ be supplier $i$'s report. We may
assume without loss of generality that any given demander finds all the
suppliers
acceptable: if demander $j$ finds supplier $i$ unacceptable, then supplier $i$
cannot have a link to demander $j$ regardless of his report, so clearly $i$'s
manipulation opportunities are more restricted. Let $\phi$ and $\phi^{'}$ be
(any) egalitarian
flows when the suppliers report $A$ and $A'$ respectively, and let $x$ and $x'$
be
the corresponding allocation to the suppliers. We show that no coalition of
suppliers
can weakly benefit by misreporting their links unless each supplier in the
coalition
gets exactly their egalitarian allocation. 

The proof is by induction on the number of type 2 breakpoints in the algorithm
to compute
the egalitarian allocation. Suppose the given instance has $n$ type 2
breakpoints, and
suppose $X_1, X_2, \ldots, X_n$ are the corresponding bottleneck sets of
suppliers.
If $n = 0$, every supplier is at his peak value in the egalitarian allocation,
and
clearly this allocation cannot be improved. Suppose $n \geq 1$.
Define $$\tilde{X}_{\ell} = \{ i \in X_{\ell} \mid \sum_{j \in A_i} \phi'_{ij}
\; \geq \;
				\sum_{j \in A_i} \phi_{ij} \},$$
and
$$\hat{X}_{\ell} = \{ i \in X_{\ell} \mid \sum_{j \in A_i} \phi'_{ij} \; \leq \;
				\sum_{j \in A_i} \phi_{ij} \}.$$
We shall show, by induction on $\ell$, that for each $\ell = 1, 2, \ldots, n$:
\begin{itemize}
\item[(a)] $\phi^{'}_{ij} = 0$ for any $i \in \tilde{X}_{\ell'}$, $j \in
\cup_{i' \in X_{\ell}} A_{i'}$, $\ell' > \ell$; and
\item[(b)] $X_{\ell} \subseteq \hat{X}_{\ell}$.
\end{itemize}
The theorem follows from part~(b) above.

Any supplier $k \in X_{\ell} \setminus \tilde{X}_{\ell}$ must have
$A_k = A'_k$ as otherwise supplier $k$ is part of the deviating coalition and
does
worse. 
Consider now a supplier $i \in \tilde{X}_{\ell}$ with $x_i <
s_i$ and a 
supplier $k \in X_{\ell} \setminus \tilde{X}_{\ell}$. We have the following
chain of
inequalities:
$$\sum_{j \in A'_k} \phi^{'}_{kj} \; = \; \sum_{j \in A_k} \phi^{'}_{kj}
				\; < \; \sum_{j \in A_k} \phi_{kj} = x_k \; \leq
\; x_i 
				\; = \; \sum_{j \in A_i} \phi_{ij}
				\; \leq \; \sum_{j \in A_i} \phi^{'}_{ij}
				\; \leq \; \sum_{j \in A'_i} \phi^{'}_{ij}.$$
To see why, note that as $k \in X_{\ell} \setminus \tilde{X}_{\ell}$, the second
inequality is true by definition, and also $A_k = A'_k$ (justifying the first
equality). Also $k, i \in X_{\ell}$ and $x_i < s_i$, implies $x_k < s_i$, as
suppliers $k$
and $i$
both belong to the same bottleneck set and supplier $i$ is below his peak; this
justifies
the third inequality. The fourth and fifth inequalities follow from the fact
that
$i \in \tilde{X}_{\ell}$ and the fact that $\phi^{'}_{ij}$ must be zero for all
$j \in A_i \setminus A'_i$.
This chain of inequalities implies that $x'_k < x_k \leq s_k$ and $x'_k <
x'_i$. 
Therefore, when the suppliers report $A'$, supplier $k$ must be a member of 
an ``earlier'' bottleneck set than supplier $i$. An immediate consequence is
that
demanders in $A'_k = A_k$ do not receive any flow from supplier $i$ when the
report is $A'$. 

By the induction hypothesis, supplier $i \in X_{\ell}$ does not send any
flow to the demanders in \textbf{$\cup_{1 \leq i' \leq \ell-1} \cup_{k \in
X_{i^{'}} } A_{k'}$.}
Therefore 
$$\{ j \mid \phi^{'}_{ij} > 0, j \in A_i\} \subseteq \{ j \mid \phi_{ij} > 0, j
\in A_i \}.$$
This observation, along with the fact that every $i \in \tilde{X}_{\ell}$ weakly
improves,
and the fact that $X_{\ell}$ is a type 2 breakpoint implies that 
$\sum_{j \in A_i} \phi^{'}_{ij} = \sum_{j \in A_i} \phi_{ij}$,
establishing~(b). Furthermore, in such a solution,
every demander $j \in A_i$ for $i \in \tilde{X}_{\ell}$ must receive all his
flow from the suppliers in $\tilde{X}_{\ell}$.In particular, the demanders in
$X_{\ell}$ 
cannot receive
any flow from suppliers in $X_{\ell'}$ for $\ell' > \ell$, establishing~(a).
To complete the proof we need to establish the basis for the induction proof,
i.e., the case of $\ell = 1$. This, however, follows easily: it is easy to
verify that the set $X_1 \setminus \tilde{X}_1$ must be empty, so 
$X_1 = \tilde{X}_1$. As $X_1$ is a type 2 bottleneck set, it is not possible
for {\em every} member of $X_1$ to do weakly better unless the allocation
remains unchanged. Thus, both (a) and (b) follow.
\qed

An easy implication is the following result, whose proof is an immediate consequence
of the results we have already established.

\begin{theorem}
In the two sided model, the egalitarian mechanism is group strategyproof
w.r.t. to both links and peaks  when the coalition is restricted to the set of
suppliers only (demanders only).
\end{theorem}

\section{Conclusions}
\vspace*{-0.2in}
\begin{figure}[h]
\begin{center}
\includegraphics[width=100mm,height=50mm]{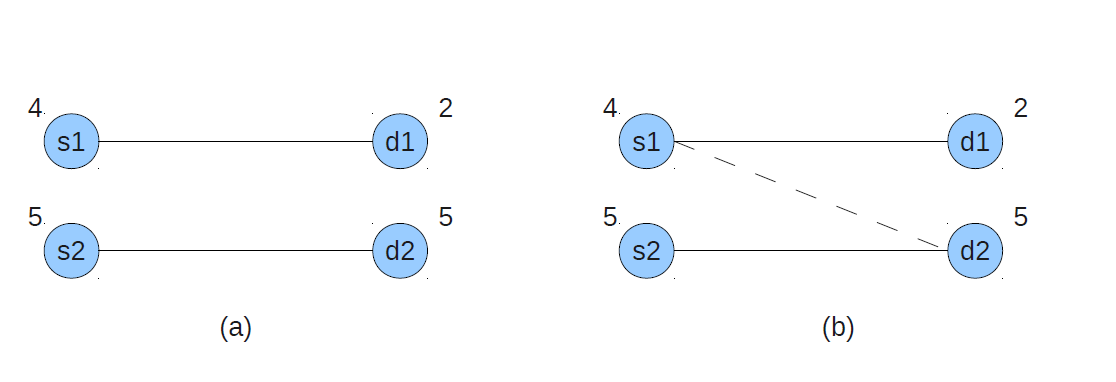}
\vspace*{-0.2in}
\caption{Bossiness of the egalitarian mechanism}
\label{fig:bossy}
\end{center}
\end{figure}
Our main contribution is a proof of peak group strategyproofness of the
egalitarian mechanism in the two settings considered by Bochet et al.
A number of interesting open problems remain, and we mention a few.
First, Bochet et al. characterize the egalitarian mechanism using
Pareto efficiency, strategyproofness, and an equity property that
can be thought of as the equal treatment of equals adapted to this
constrained setting. A natural question is if there is an alternative
characterization that uses group strategyproofness, but weakens the
efficiency or the equity requirement. Second, it will be interesting
to characterize all (both link and peak) strategyproof mechanisms 
for this problem. In particular, it will be of interest to find natural
link-monotonic mechanisms for the one-sided model (the egalitarian
mechanism fails this test). Finally, note that the egalitarian mechanism
has the somewhat undesirable property that an agent can change the
allocation of other agents without altering his own, see the example
of Figure~\ref{fig:bossy}. (Such mechanisms are said to be ``bossy.'')
Are there natural non-bossy mechanisms for
this problem that are compelling from a normative point of view?

\newpage
\nocite{*}
\bibliographystyle{plain}
\bibliography{refs}
\newpage

\end{document}